# Multiband NFC for High-Throughput Wireless Computer Vision Sensor Network

Fei Y. Li, Jason Y. Du
09212020027@fudan.edu.cn

Vision sensors lie in the heart of computer vision. In many computer vision applications, such as AR/VR [1], non-contacting near-field communication (NFC) with high throughput is required to transfer information to algorithms. In this work, we proposed a novel NFC system which utilizes multiple frequency bands to achieve high throughput.

1. **Computer Vision and Sensors**

With emerging machine learning algorithms [1-5], computer vision tasks such as face recognition, object detection and simultaneous localization and mapping (SLAM) are now using more and more machine learning methods. In order to achieve good results, large amount of data is required to feed into algorithm. This not only requires energy-efficient computation [6-7], but energy-efficient data transfer is also necessary.

Point-to-point file exchange among smart devices has attracted people's attention today. The smart devices support high-definition photo/video capture (some smart phones even support 4K definition photo/video capture), and many people are actually using smart devices to replace camera. As 4K photos and videos have large file size (>10 MB per photo and >300 MB for video per minute), the most convenient way to share those photos with friends is by high-speed point-to-point link. In addition to file transfer, many computer vision application categories such as embedded systems and military applications also desire to have vision sensors coupled to computation unit wirelessly. Therefore, the connection/coupling between sensor and processor can be more flexible and more reliable. The distance between sensor and processor does not necessarily to be long, and near-field communication is a good match for this case (Fig. 1).

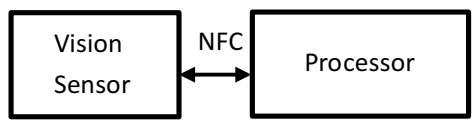

Fig. 1 Processor and vision sensor coupled by NFC

However, data rate of NFC is usually low. Currently several standards and products have been proposed for high-speed point-to-point communication. Bluetooth 4.0 has one HS mode, which uses ad-hoc point-to-point WiFi link to transfer data. The highest data rate can reach 25 Mbps. WiFi Alliance also proposed similar WiFi Direct technology, which supports data rate as high as 250 Mbps. However, the protocols of Bluetooth and WiFi both require a long time (can be as long as >10 seconds) to establish link (searching and pairing), and the user experience is not very smooth. In addition the bandwidth of WiFi and Bluetooth is not scalable due to spectrum regulation, and the data rate upgrade of Bluetooth/WiFi is much slower than the growth of

computer vision requirements

## 2. Multiband RF Interconnect for NFC

High-speed NFC has the potential to solve above issues. Per FCC regulation, NFC must work in several pre-defined license-free industry, science and medic (ISM) bands, such as 900 MHz, 2.4 GHz, 5.8 GHz, etc [8]. Each frequency band only has limited bandwidth, so the conventional NFC which only uses one frequency band has limited data rate. On the other hand, NFC limits communication range to <3 cm, and therefore output power of TX can be low enough to pass spectrum regulation. This means that NFC can use broader bandwidth for data transfer. In addition, high speed NFC can use a simple protocol for link. By selecting frequency band properly, we can make sure there is no interference from other communications. There is only one TX-RX pair within communication range, and protocol can be simplified. This means the link establishment can be very quick, and user experience for data sharing will be very smooth. The diagram using high-speed NFC is shown in Fig. 2. A designed coupler is placed between sensor and processor, and the other part of device is shielded to reduce RF power leakage.

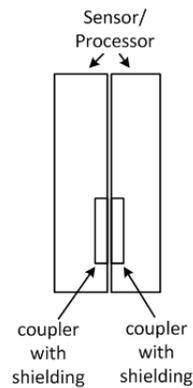

Fig. 2 Processor and vision sensor interface with coupler/shielding.

On the other hand, multiband RF signaling can be used to enhance data rate. By divide data stream into multiple sub-streams and each up-converted to one ISM band, multiple data streams can be transferred simultaneously. The multiple RF signals can be combined by power combiner [9].

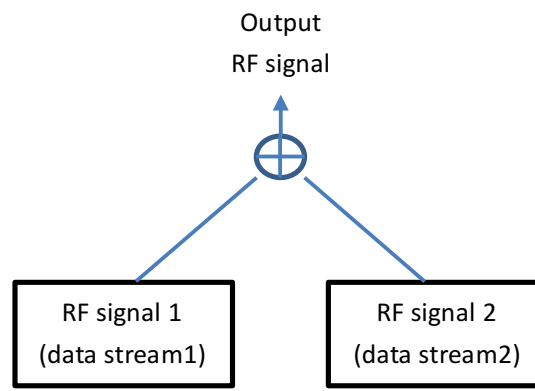

Fig. 3 Multi-band NFC

## 3. All-Digital Transmitter (ADTX) with FPGA

An all-digital transmitter design methodology has been proposed by Li et al. in [10]. In [10], transmitter is implemented by synthesis, therefore greatly reduces the turnaround time of transmitter design. It is an ideal solution to our prototyping of high-speed NFC.

We used the same transmitter architecture as in [10], using SDM and XOR mixing to convert analog signal into high-speed RF signal. The transmitter architecture is shown in Fig. 4.

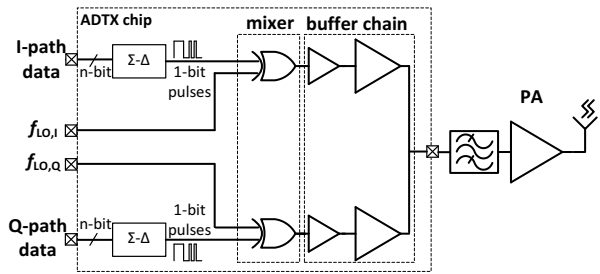

Fig. 4 ADTX architecture

Further, we used FPGA to fast implement ADTX circuit. Also, due to flexible reconfiguration of FPGA, we can repeatedly iterate our design until design goal is met.

## 4. Implementation Results

We implemented our high-speed NFC system, including one coupler and one FPGA-based ADTX. The coupler is shown in Fig. 5, and its coupling loss is around 10-20 dB for near-field (Fig. 6).

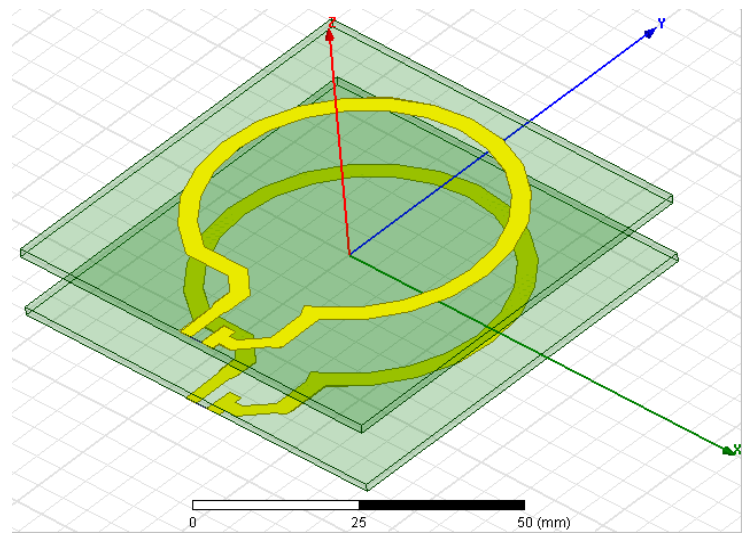

Fig. 5. Coupler design.

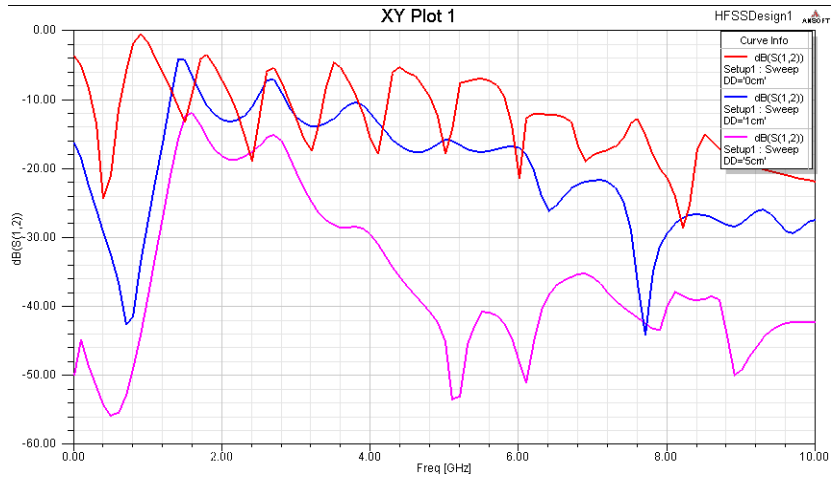

Fig. 6 Simulation results of coupling loss

The output spectrum of NFC system is captured by spectrum analyzer, as shown in Fig. 7. We used 250MHz and 400 MHz as carrier bands, and two peaks appear in spectrum analyzer.

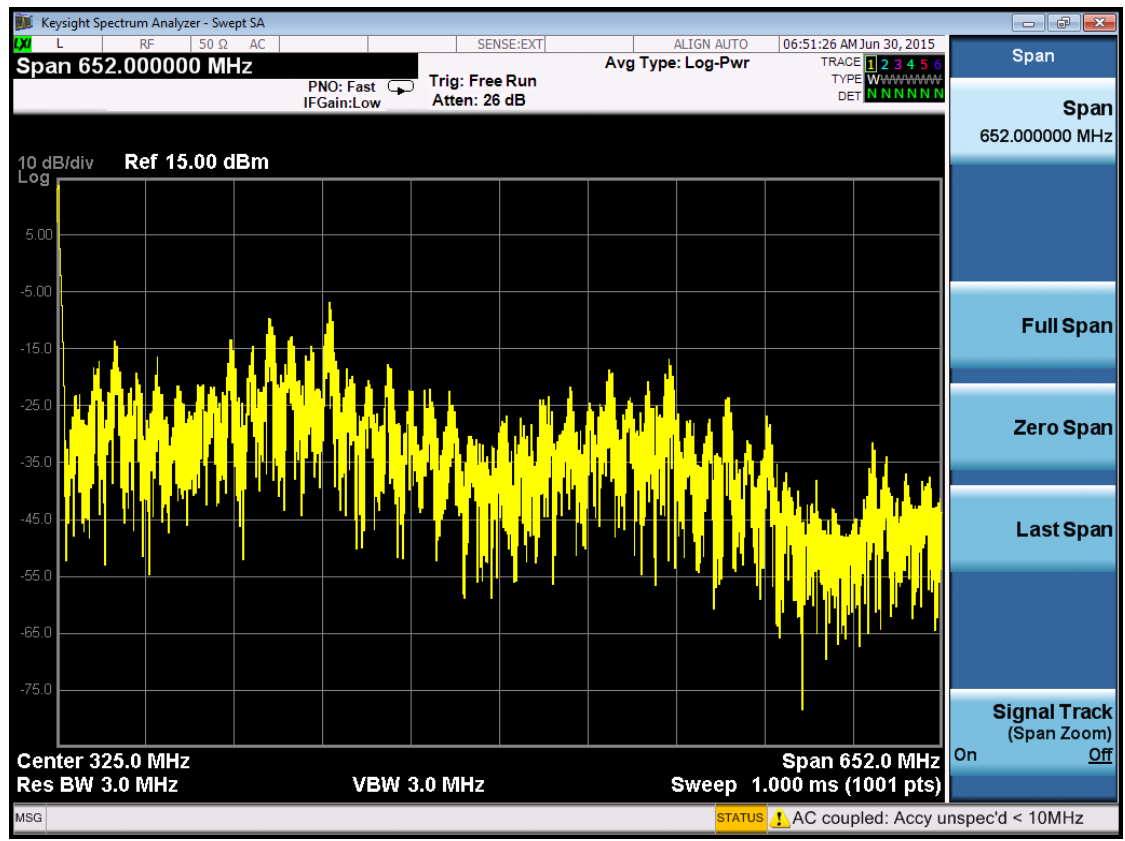

Fig. 7 Spectrum of TX output

The demodulated waveform is shown in Fig. 8. QAM-16 is used for modulation, and it can be demodulated successfully. The data rate per band is 16 Mbps, and total data rate is 32 Mbps. The data rate is limited by carrier frequency, and carrier frequency is limited by speed of FPGA. With ASIC implementation, we expect carrier frequency as high as 6 GHz [10], as total data rate of greater than 300 Mbps can be achieved, which can meet the requirement of computer vision.

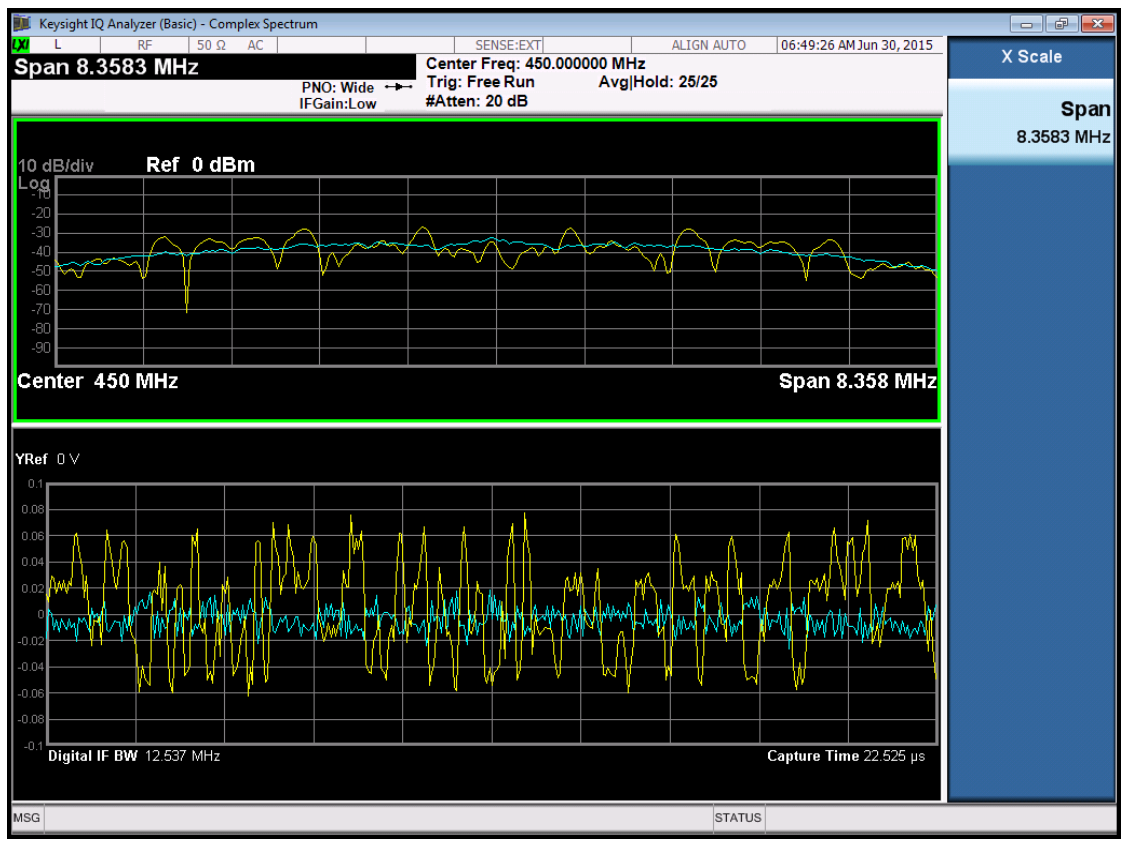

Fig. 8 Demodulated QAM-16 data at 450 MHz.

**Conclusion**

In this work, we proposed a high-speed NFC system for computer vision applications. This system includes coupler and ADTX to achieve high speed data transfer, while achieves faster prototyping for non-contact data transfer. The data rate can be further boosted with ASIC implementation.